\font\biggtenrm=cmr10 scaled\magstep2

\font\biggtenbf=cmbx10 scaled\magstep2

\newcommand{\RM}{\rm}

\def\beq{\begin{equation}}
\def\eeq{\end{equation}}

\def\beqy{\begin{eqnarray}}
\def\eeqy{\end{eqnarray}}

\input epsf
\documentstyle[12pt,aaspp4]{article}
\begin{document}
\begin{titlepage}
\centerline{\biggtenbf Analytical Dynamical Models for Double-Power-Law Galactic Nuclei}
\vskip 1cm
\centerline{\biggtenrm  HongSheng Zhao }
\centerline{Max-Planck-Institute f{\"u}r Astrophysik, 85740 Garching, Germany}
\centerline{Email: hsz@MPA-Garching.MPG.DE}
\vskip 0.5cm
\centerline{\biggtenbf ABSTRACT}

Motivated by the finding that the observed surface brightness profile
of many galactic nuclei are well fit by double-power-laws, we explore
a range of spherical self-consistent dynamical models with this light
profile.

We find that the corresponding deprojected volume density profile,
phase space density and line-of-sight velocity distribution of these
models are well fit by simple analytic approximations.  We illustrate
the application of these models by fitting a sample of about 25
galactic nuclei observed by Hubble Space Telescope.  We give the
derived volume density, phase space density, velocity dispersion and
line profile parameters in tables.  The models are useful for
predicting kinematic properties of these galaxies for comparison with
future observations.  They can also be easily applied to seed N-body
simulations of galactic nuclei with realistic density profiles for
studying the evolution of these systems.

\vskip 0.5cm
\centerline{{\it Subject headings}: galaxies: kinematics and dynamics - line: profiles }
\vskip 3cm
\centerline{Submitted to Monthly Notices of R.A.S.}

\end{titlepage}


\section{Introduction}
Recent high-resolution observations of galactic nuclei find that their
surface brightness profiles are well fit by a three-parameter
double-power-law.  Assuming a constant mass-to-light ratio, this means
that the projected mass density $\mu(R)$ satisfies (Lauer et al. 1995,
Byun et al. 1996)
\footnote{
The parameters $\alpha_1,\beta_1,\gamma_1$ in this paper correspond to
${1 \over \alpha}, \beta, \gamma$ respectively in the convention used
by these authors.  Their models are restricted to surface brightness
studies to establish the central cusp in light.}
\begin{equation}\label{mu}
\mu(R) = \mu_0\left({R \over B}\right)^{-\gamma_1} 
\left(1 +\left({R \over B}\right)^{1 \over \alpha_1}\right)^{-(\beta_1-\gamma_1)\alpha_1},
\end{equation}
where $\gamma_1$, $\beta_1$ and $\alpha_1$ specify the slope of the
inner power-law, the outer power-law and the width of transition
region.  Meaningful values for the three parameters satisfy $0\leq
\gamma_1 \leq \beta_1$ and $\alpha_1>0$.  Other parameters $B$, $\mu_0$
are scaling parameters, specifying a length scale (the break radius)
and a scale for projected density.  For flattened nuclei, the
parametrization applies on the major or minor axis or in a shell
averaged sense.

Interpreting this class of density profiles requires a new class of
dynamical models.  Previous dynamical models mostly have a finite
core.  A recent set of cuspy models, including the $\gamma/\eta$
models discovered by Dehnen (1994) and Tremaine et al. (1995), and a
even wider range of analytical models by Zhao (1996), often do not
match the light of nuclei outside the break radius.  Unlike these
models the observed light profiles are often much shallower at large
radius, and in principle correspond to a divergent mass if
extrapolated to infinity.  Hence it is necessary to explore dynamical
models consistent with a general double-power-law profile, including
those with a divergent mass.

The existence of a universal parametrization constitutes a major
advantage for theoretical study of their dynamical properties.  It
makes it possible to study observed galactic nuclei as a class spanned
by the three slope parameters $(\alpha_1, \beta_1, \gamma_1)$, and
constrain the models with the steady state dynamics without
necessarily using data of individual systems explicitly.  In this
paper, we give some results which immediately follow from the above
surface brightness profiles.  We will concentrate on the simple class
of spherical isotropic models with main emphasis on their simple
analytical results.

While spherical models with a $f(E)$ distribution function are famous
for admitting mathematical solutions, and are often used as a
compromise for more realistic but less tractable
anisotropic/flattened/triaxial models, the actual implementation of
the spherical models is in fact very tedious, and rarely admits simple
analytical results.  These significantly complicate the first-level
simple interpretation of photometric and kinematic data.  The standard
inversion using Eddington formula involves at least three integrations
and two derivatives (analytical or numerical) to get $f(E)$ from a
surface density profile.  To further make a prediction on line profile
requires computing a three dimensional integral (see e.g., Dehnen
1994) at each grid point in the projected radius vs. velocity plane.

For the current problem, one is interested in a class of models with a
range of density profiles.  The main challenge is to present the
results in a manageable and easily interpretable way.  Virtually no
rigorous analytical solutions are known for the projected models given
by Eq.(1).  Even for some related analytical models (Zhao 1996, Dehnen
1994, Tremaine et al. 1995), the expressions of the distribution
function and projected density and dispersion are generally very
lengthy, typically involving more than half a dozen analytic terms
with no clear physical meaning and with possible large cancellations
between terms.  As a result, the relations between observable and
model quantities are obscured.

In this paper we build a set of spherical dynamical models with simple
functional forms for the intrinsic volume density and $f(E)$.  We fit
these to the double-power-law projected density models.  We set the
scaling quantities, $\mu_0$ and $B$ as unity, and vary the
dimensionless parameters $(\alpha_1, \beta_1, \gamma_1)$ in a 3D
parameter space to simulate a complete set of radial profiles of the
surface light.  Because of the way the fitted models are tailored, the
residuals of the fits are typically smaller than the uncertainties in
the data so that the models are practically consistent with the
double-power-law surface brightness profile.

The main results of the paper are summarized in several universal
formulae for the volume density, the phase space density, the
line-of-sight velocity profiles.  The model results can be presented
directly with the fitting formulae and a few numerically fitted
parameters.

To demonstrate the applications, we compute the system parameters for
about 25 observed galaxies and give them in tables.  With these the
intrinsic and projected quantities of the model are fully determined,
and it involves virtually no further calculation to predict the
line-of-sight velocity distributions.

The paper is organized as follows.  In Section 2 we fit the surface
brightness profiles with analytical volume density models.  Section 3
gives the analytical expression for the model potential.  Section 4
gives the deprojected phase space density by matching the volume
density and the potential.  In Section 5, the distribution functions
are re-projected to yield line-of-sight velocity distributions as well
as the dispersion and kurtosis of the profiles, all on a grid of
projected radius.  We illustrate the model applications in Section 6.
We summarize in Section 7.  Asymptotic relations for the model
quantities are given in Appendix A.  Some alternative analytical
approximations with rigorous asymptotic solutions are given in
Appendix B.  A simple formula for the line profiles of the models is
derived in Appendix C.

Although similar models can be built for anisotropic spherical systems
with a black hole, and the techniques are also generalizable to oblate
$f(E,J_z)$ systems, we leave these generalized models for a later
study (Zhao and Syer 1996).  The three-dimensional $(\alpha_1,\beta_1,
\gamma_1)$ parameter space of the spherical models is already very
big, and at least two more dimensions would be necessary to cover
spherical models with black hole and anisotropy.  Also while the
isotropic models are most likely stable, many simulations are needed
to examine the stability of anisotropic or black hole models before
applying them to observations.

\section{Deprojected volume density profile}

One expects an asymptotic power-law in surface brightness to
correspond to a (steeper) power-law in volume density.  So the
double-power-law of the observed surface brightness of galactic nuclei
suggests that their volume density could be fit by a similar
double-power-law,

\begin{equation}\label{nu}
\nu(r) = \nu_0 \left({r \over b}\right)^{-\gamma} 
\left(1 +\left({r \over b}\right)^{1 \over
\alpha}\right)^{-(\beta-\gamma)\alpha},
\end{equation}
where the new parameters $(\alpha, \beta, \gamma)$ and $\nu_0$ and $b$
have the same meaning as the five parameters in Eq.~\ref{mu} except
that they describe the volume density profile instead of the surface
brightness profile.  In Appendix A we give the relations between
$(\alpha, \beta,\gamma, \nu_0, b)$ and $(\alpha_1, \beta_1,\gamma_1,
\mu_0, B)$ based on matching the two densities at asymptotic large or small
radii.  But except for a few special cases\footnote{for finite core
models with $\alpha=\alpha_1={1 \over 2}$ and for single-power-law
models}, the deprojected density of $\mu(R)$ in Eq.~\ref{mu} is
generally not as simple as $\nu(r)$ in Eq.~\ref{nu}.

Our approach is to match the two densities in the least squared sense.
We fix the outer slope
\begin{equation}
\beta=\beta_1+1,
\end{equation}
to enforce a good match at large radius.  Then we tune the other four
parameters $\alpha, \gamma, \nu_0, b$ to fit the surface brightness to
a high accuracy by minimizing the following r.m.s residual,
\begin{equation}\label{ressurf}
\delta_\mu = \left({1 \over N_R} \sum_{i=0}^{N_R} [\log I(R_i) - \log \mu(R_i)]^2\right)^{1 \over 2}
\end{equation}
where
\begin{equation}\label{IR}
I(R) =2\int_0^{+\infty} \nu(\sqrt{R^2+z^2}) dz
\end{equation}
is the projected density of $\nu(r)$, and the fitting positions $R_i$
and the number of fitting points $N$ are given in Table 1.

The initial values for the fitting parameters are taken from 
the following approximate relations
\begin{eqnarray}
\gamma \approx \gamma_{e} &= & \gamma_1+1 \mbox{ \RM{if $\gamma_1 >0$} } \\
       & = & [1-{1 \over \alpha_1},0]_{max} \mbox{ \RM{if $\gamma_1 =0$  } } 
\end{eqnarray}
to fit near the cusp, where $\gamma_{e}$ is the expected exact
asymptotic power for the density, and
\begin{equation}
\alpha \approx \alpha_1, \,\,\, b \sim B,
\end{equation}
to fit near the break radius, and
\begin{equation}
 \nu_0 b^{-\beta} \sim 
{\Gamma({\beta \over 2}) \over 2\sqrt{\pi} \Gamma({\beta - 1 \over 2}) }
\mu_0 B^{-\beta-1} ,
\end{equation}
to match the density normalization far away.

The first panel in Fig. 1, Fig. 3 and Fig. 5 shows a few typical model
fits and the upper panel of Fig. 6 shows their residuals.  The
residual is small over 4 or more decades in the intensity scale.  The
typical residual $|\log I(R)- \log \mu(R)| \sim \delta_\mu$ is
$10^{-4}-10^{-1}$ for the projected radius $0.01 \le R/B \le 100$.
For galactic nuclei observed by Space Telescope the light is measured
to an accuracy of about $0.1$ magnitude ($\sim$ 4\% difference in ten
log) from the central $0.1\arcsec$ to $10\arcsec$ outward, with a
typical break radius at $1\arcsec$.  So the small internal residuals
of the models here are negligible when fitting the photometric data.
The above results are independent of the dynamics.

\section{The potential}

Zhao (1996) shows that the potential $\phi(r)=-\Phi(r)$ corresponding to a
double-power-law volume density distribution $\nu(r)$ is simply a sum
of two (analytical) incomplete Beta-functions (Zhao 1996).  
\begin{equation}\label{phi}
\Phi(r)  =  4\pi \nu_0 b^2 \alpha [ {b \over r} 
B(\alpha ( 3-\gamma ), \alpha ( \beta -3), 
{({r \over b})^{1 \over \alpha} \over 1+ ({r \over b})^{1 \over \alpha} })
 + B(\alpha ( \beta -2), \alpha (2 -\gamma), 
{1 \over 1+ ({r \over b})^{1 \over \alpha} }) ],
\end{equation}
where the gravitational constant $G$ is set to unity. 

As the incomplete Beta-functions can be computed with a fast function
call to standard routines in Numerical Recipes (Press et al. 1992),
this greatly simplifies the numerics for the dynamics.

According to the asymptotic expression for $\Phi(r)$ given in Appendix
A, the zero point of the potential is at infinitely large radius.  The
depth of the potential well $\Phi_0=\Phi(0)>0$ is infinite for models
with a strong cusp with $\gamma \geq 2$ (or if there is a central
black hole), and is finite for $\gamma<2$.

\section{Deprojected distribution function}

In this section, we derive an approximation to the underlying
intrinsic distribution function of the double-power law spherical
density models.  For simplicity we consider only models with isotropic
velocity distribution, namely, models with distribution function
$f=f(Q)$, where we define a positive energy $0 \leq Q \leq
\Phi(0) $ with
\begin{equation}\label{Qdef}
Q \equiv -E =\Phi(r) -{1 \over 2} v^2,
\end{equation}
and a function $G(Q)$ being an integration of the $f(Q)$,
\begin{equation}\label{gqdef}
G(Q) \equiv \int_0^Q f(Q) dQ.
\end{equation}

To decide a functional form for $f$ or $G(Q)$, we note that the
distribution function of asymptotic power-law systems is often a
power-law of energy at asymptoticly large or small radius.  For
example, in the Hernquist model, $f(Q) \propto Q^{5/2}$ at large
radius and $f(Q) \propto (\Phi_0-Q)^{5/2}$ at small radius.  So a
sensible universal formula for the distribution function of the
double-power-law density models should be a smooth positive function
of $Q$ which reduces to a power-law  at small $Q$ and large
$Q$.

The following contrived expression for $G(Q)$ has the desired property
\begin{equation}\label{intdf}
G(Q) = f_0 Q_b q^{\beta_2} 
\left(1+q^{1 \over \alpha_2}\right)^{(\gamma_2-\beta_2)\alpha_2},
\end{equation}
where $f_0$ and $Q_b$ are two scaling quantities with the
dimension of phase density and energy, $(\alpha_2, \beta_2, \gamma_2)$
are another three dimensionless quantities for the shape of the
distribution.  We define a dimensionless energy $q$, which is a
rescaling of the energy $Q$ with
\begin{eqnarray}\label{qdef}
q & \equiv &{ {Q \over Q_b} \over 1 - {Q \over \Phi_0 } }, 
\mbox{ \RM{for finite potential well,}} \\
  & \equiv &  {Q \over Q_b}, \mbox{ \RM{for infinitely deep potential well,}} 
\end{eqnarray}
so that $q$ runs from $0$ to $\infty$ with decreasing radius for
models with finite as well as infinitely deep potential well.  Note
smaller (bigger) values for $Q$ or $q$ correspond to bigger (smaller)
radius, and $G(Q)$ reduces to power-law of $q$ at big $q$ and at small
$q$.

Taking the derivative of Eq.~\ref{intdf} we obtain the corresponding
distribution function $f(Q)$ as
\begin{equation}\label{df}
f = f(Q) = f_0 q^{\beta_2-1} 
(1+q^{1 \over \alpha_2})^{(\gamma_2-\beta_2)\alpha_2} \times W \times U,
\end{equation}
where
\begin{equation}\label{wdef}
W \equiv \gamma_2+ {\beta_2-\gamma_2 \over 1+q^{1 \over \alpha_2} },
\end{equation}
and 
\begin{eqnarray}\label{udef}
U & \equiv  &{ 1 \over (1 - {Q \over \Phi_0})^2 } = 
(1+  q{Q_b \over \Phi_0 })^2 , 
\mbox{ \RM{for finite potential well,}} \\
  & \equiv  &  1, \mbox{ \RM{for infinitely deep potential well,}} 
\end{eqnarray}
are two dimensionless factors.

It then follows from the above equation that $f(Q)$ is positive
definite given that $\gamma_2 \ge 0$ and $\beta_2 \ge 0$, and has the
following asymptotic power-law dependence on $Q$,
\begin{eqnarray}\label{fasy}
f(Q) & \propto & Q^{\beta_2-1}, \mbox{ \RM{ $Q \rightarrow 0$}}\\
     & \propto & Q^{\gamma_2-1}, \mbox{ \RM{ $ Q \rightarrow \Phi_0=\infty $}} \\
     & \propto & (\Phi_0 - Q)^{-\gamma_2-1}, \mbox{ \RM{ $ Q \rightarrow $ a finite $ \Phi_0 $ and $\gamma_2>0$ }} \\
     & \propto & 1, \mbox{ \RM{ $ Q \rightarrow $ a finite $ \Phi_0 $ and $\gamma_2=0$ }}.
\end{eqnarray}

To briefly comment on more general models, the parametrization for
$G(Q)$ or $f(Q)$ is also plausible for models with a central black
hole as $f(Q)$ is a power-law for infinitely deep potential well
(Tremaine et al. 1994).  One can also obtain its simple counterpart in
Osipkov-Merritt type anisotropic models if replacing $Q$ with
$Q_a=-E-{1 \over 2} \eta J^2)$.

For a given potential $-\Phi(r)$, the distribution function $f(Q)$ has
five fitting parameters $(\alpha_2, \beta_2, \gamma_2, Q_b, f_0)$.
These are determined by making $f(Q)$ and the volume density $\nu(r)$
consistent.  In practice we minimize the following r.m.s. residual,

\begin{equation}\label{resden}
\delta_\nu = \left({1 \over N_r} \sum_{i=0}^{N_r} [\log n(r_i) - \log \nu(r_i)]^2\right)^{1 \over 2}
\end{equation}
where the fitting positions $r_i$ and the number of fitting points
$N_r$ are given in Table 1, and
\begin{equation}\label{nr}
n(r) = 4\pi \sqrt{2} 
\int_0^{\Phi(r)} f(Q) \sqrt{\Phi(r) -Q} dQ
\end{equation}
is the volume density corresponding to $f(Q)$ (cf. Binney and Tremaine
1987); after an integration of parts it reduces to a more convenient
expression for $n(r)$:\footnote{ A useful generalization to
Osipkov-Merrit models can be obtained by specifying the distribution
function $f=f(Q_a=-E-\eta J^2)={d \over dQ_a}G_a(Q_a)$, where
$G_a(Q_a)=G(Q_a)+\eta g(Q_a)$ is a linear function $\eta$, and it
reduces to $G(Q)$ for isotropic model.  In this case, one can show
(with Eq. 4-148 of Binney and Tremaine) that $n(r) (1+\eta r^2) = 2\pi
\sqrt{2} 
\int_0^{\Phi(r)} {G(Q_a)+ \eta g(Q_a) \over \sqrt{\Phi(r) -Q_a}} dQ_a$, and
$G(Q_a)$ and $g(Q_a)$ depend on $\eta$ only through $Q_a$.}
\begin{equation}\label{nr1}
n(r) = 2\pi \sqrt{2} 
\int_0^{\Phi(r)} {G(Q) \over \sqrt{\Phi(r) -Q}} dQ.
\end{equation}

There are several simple (approximate) relations between the
parameters $(\alpha_2, \beta_2, \gamma_2, Q_b, f_0)$ and $(\alpha,
\beta, \gamma, b, \nu_0)$, which follows from matching the densities
$\nu(r)$ and $n(r)$ at big and small radius.  If $\beta_{2e}$ and
$\gamma_{2e}$ are the exact asymptotic powers for $G(Q)$ as given in
Apppendix A, we require $\beta_2 =\beta_{2e}$ in the fitting program.
The true fitting parameters reduce to only four.  The somewhat rigid
form of the five-parameter distribution function $f(Q)$ prevents
fixing $\gamma_2$ to $\gamma_{2e}$: in order to fit everywhere about
equally well, the fitted function does not follow exactly the
analytical asymptotic behavior.  Alternative solutions to cure this
problem are discussed in Appendix B.

Some fits are shown in the lower left panels of Fig. 1, Fig.3 and
Fig. 5, and their residuals are shown in the lower panel of Fig.6.
The reprojected volume density $n(r)$ can often fit $\nu(r)$
satisfactorily for several decades in density with typical residuals
$|\log n(r) - \log \nu(r)| \sim \delta_\nu$ from 0.1 to 0.001 for radius
$ 0.01 \le r/B \le 100$.  
Table 5 gives the parameters of the models shown in Fig. 1, Fig. 3 and Fig. 5.

\section{Re-projected velocity profiles}

The projected velocity profiles are the main observable constraints to the
dynamical mass distribution of a system.  Since the profiles $L(R,
v_z)$ are functions of both projected radius $R$ and line-of-sight
velocity $v_z$, one often prefers to use moments of the profiles at
selected projected radius as convenient comparisons with observation.
For the spherical $f(E)$ models here, the odd moments such as rotation
and skewness are all zero, and most of the information of the profiles
are contained in the lowest order even moments, namely, the line
intensity (presumably proportional to the projected density
$\mu(R)$), the dispersion $\sigma(R)$ and the kurtosis.

There are several ways to represent a profile with the lowest moments.
While the Gauss-Hermit expansion (van der Marel and Franx 1993) is
mathematically elegant, Zhao and Prada (1996) showed that the direct
Gauss-Hermit expansion genericly gives rises to profiles with negative
wings and multiple peaks.  To cure these problems without losing the
elegance and many nice properties of the Gauss-Hermit expansion, they
proposed the following fitting formula for line profiles, which we
will use.

\begin{equation} \label{linefit}
L (R, v_z) = {L_0(R) \over \sqrt{2 \pi} \sigma} 
e^{-{v_z^2 \over 2 \sigma(R)^2}}
\{ 1  + \lambda e^{-{ v_z^2 \over 2 \sigma(R)^2}} 
c_4(R) H_4 (\lambda { v_z \over \sigma(R) } ) \},~~~ \lambda=\sqrt{3 \over 2},
\end{equation}
where $H_4(y)$ is the four-order Hermit polynomial of $y$
\begin{equation}
H_4 (y)= {2 \over \sqrt{6}} (y^4 - 2y^2 + {3 \over 4}),
\end{equation}
and $\sigma(R)$ is the best-fit dispersion at radius $R$, $c_4(R)$ is
a parameter describing the kurtosis of the profile.  This fitting
formula differs mainly from the usual Gaussian-Hermit expansion by the
extra Gaussian damping term $\sqrt{3 \over 2} e^{-{ v_z^2 \over 2
\sigma(R)^2}}$ in front of the Hermit polynomial $H_4$, which helps to
suppress oscillatory peaks far from systematic velocity and to
eliminate the unphysical negative wings; the coefficients $\sqrt{3
\over 2}$ in the damping term and in the $H_4$ are to keep the
orthogonality of the functions.  The formula is robust for mildly
double-peaked $-0.25 \le c_4 \le -0.15$ or mildly cuspy $0.2 \le c_4
\le 0.45$ profiles and for profiles close to Gaussian $-0.1 \le c_4 \le 0.1$.
The conventional $h_4$ parameter is approximately $c_4$ for nearly
Gaussian profiles, but the former cannot fit mildly non-Gaussian
profiles.

For the dynamical models here, $\sigma(R)$ and $c_4(R)$ are determined
by fitting $L(R, v_z)$ to the projected velocity distribution $P(R,
v_z)$ at each projected radius $R$.  We minimize the 
following r.m.s. residual at each radius $R$,

\begin{equation}\label{chiline}
 \left({1 \over N_v} \sum_{j=0}^{N_v} [L(R, v_j) - P(R, v_j)]^2\right)^{1 \over 2},
\end{equation}
where the velocity grid $v_j$ and the number of points $N_v$ is given
in Table 1; the velocities $v_j$ scales with the escape velocity at
positions $R_i$.

The projected velocity distribution $P(R, v_z)$ at projected radius
$R$ of the $f(Q)$ models is simply
\begin{eqnarray}\label{profile}
P(R, v_z) & \equiv & \int^\infty_{-\infty} dz \int\int dv_x dv_y f(Q) \\
          & = & 4\pi \int_0^\infty dz 
{G(Q=\Phi(\sqrt{R^2+z^2}) + 
         {1 \over 2} v_z^2)  },
\end{eqnarray}
where $G(Q)$ is defined in Eq.~\ref{intdf}.  So the profile $P(R,
v_z)$ has been reduced from a 3D to a 1D line-of-sight integration.  See
Appendix C for derivation of this equation and its generalized form for
anisotropic systems.

Fig. 2 and 4 show some typical line-of-sight velocity distribution
fits at three different radii $R=0.1,1,10$.  One can see that the
fitting formula can recover the profiles $P(R, v_z)$ to good accuracy.
The residual is typically between $0.01$ and $0.05$ of the peak
intensity.

The right panels of Fig.1, Fig.3, and Fig. 5 also show the radial run
of the dispersion and kurtosis for a few models.  As noted in Tremaine
et al. (1994), depending on the strength of the central cusp, the
dispersion can have a peak near the break radius (if $1 \le \gamma <
2$) or a steady falling radial profile (if $\gamma \ge 2$ or $\gamma <
1$).  

\section{An application to observed galaxies}

As an illustration of the models, we apply them to a sample of observed
galactic nuclei given in Byun et al (1996).  The intrinsic volume
density and phase space density parameters of these systems are
derived and listed in Table 4.  

Interestingly most of observed nuclei have a divergent total mass if
the light profile is extrapolated to infinity because their
$\beta=\beta_1+1 \le 3$.  The typical values for $\alpha_1 \sim 0.5$.
These properties would not be adequately accounted for by a known
narrow class of analytical models by Dehnen (1993) and Tremaine et
al. (1994), which are characterized with $\beta=4$ and $\alpha=1$.

The residuals of our proposed analytical models for the volume density
and phase space density of galactic nuclei are about as small as (if
not smaller than) the residual for the double-power-law used in Byun
et al. fitting the photometric data.  We conclude that the universal
surface brightness profile also corresponds to a universal volume
density and phase space density.  And the analytical models here are
secure for interpreting observation.

Given this, we can make predictions on observable kinematics.  The
fitted values of $\sigma(R)$ and $c_4(R)$ on a radial grid are given
in Table 5.  To obtain values at other radius, one can simply
interpolate between the tabulated values as both quantities are smooth
functions of the radius.

As it is clear from Table 5, the model predicts a very small kurtosis
near or outside the break radius of these observed nuclei.  We note
that this is general as shown in the right bottom panels of
Fig. 1, Fig. 3 and Fig. 5 for hypothetic systems.  We find that for
the whole class of double-power-law isotropic models with $\le 0
\gamma_1 < 2$ and $\beta_1 \ge [2,\gamma_1]_{max}$ and $0.5 \le \alpha_1
\le 2$, the profiles are always very close to Gaussian with $-0.05
<c_4(R) <0.2$.  The amplitude of $c_4(R)$ generally increases towards
the center, but is small at all radii.  Outside the core, $R \ge 1$,
the kurtosis is negligiblely small with $|c_4(R)| \le 0.03$.  These
results argue that velocity profiles are always very close to Gaussian
for the whole class of isotropic double-power-law models.  They
support the interpretation that strongly non-Gaussian profiles near or
inside the break radius are indications of either anisotropy or
central black hole.

\section{Summary}

In summary, a large number of galactic nuclei obey a parametrized
double-power-law surface brightness radial profile (Byun et al. 1996).
We find that their intrinsic volume density fits a similar universal
double-power-law with a comparble residual.  We further explore
spherical isotropic models consistent with these profiles, and find a
simple fitting formula for the distribution function $f(E)$ as well.
These parametrizations are tailored so that their functional forms
reduce to power-laws at large or small radius.

These analytical models also simplify the procedures to interpret
photometric and kinematic data of galactic nuclei.  We demonstrate the
models with a simple application to a group of observed galactic
nuclei, and predict the radial runs of their velocity dispersion and
kurtosis.  Tables for computed models as well as FORTRAN programs to
run additional models are available at site
http://ftp.ibm-1.mpa-garching.mpg.de/pub/hsz.

Galactic nuclei are generally flattened with a possible central black
hole and velocity anisotropy.  For these models the distribution
function is generally a function of two or three integrals,
$f=f(E,J_z,I_3)$.  Still the simple spherical model here can provide
some simple insights which help to build these more complex models.
We expect that an $f(E,J_z,I_3)$ with its energy dependence similar to
the fitting formula for isotropic models here will give a plausible
fit to anisotropic flattened systems if with a double-power-law radial
profile.

I thank Dave Syer for a critical reading of the manuscript and many
helpful comments.

\vfill\eject

\section{Appendix}
\appendix

\section{Asymptotic expressions of the double-power-law models}

Here we give the asymptotic expressions for the projected density, the
phase space density and the potential for the double-power-law {\it
volume} density model.

For the volume density $\nu(r)$
\begin{eqnarray}\label{nuasm.1}
\nu (r) & \rightarrow & \nu_0 ({r \over b})^{-\beta}
\mbox{ \RM{if $r\rightarrow +\infty$}}, \\ \label{nuasm.2}
        & \rightarrow & \nu_0 ({r \over b})^{-\gamma}
\mbox{ \RM{if $ r \rightarrow 0$}},
\end{eqnarray}

For the projected density $I(R)$
\begin{eqnarray}
I(R) & = & 2\int_0^{+\infty} \nu(\sqrt{R^2+z^2}) dz \\ \label{irasm.1}
     &\rightarrow & c_{\beta-1} ({R \over b})^{1-\beta}
\mbox{ \RM{if $R \rightarrow +\infty $ }} \\ \label{irasm.2}
     &\rightarrow & I_0 H(1-\gamma)+ c_{\gamma-1} ({R \over b})^{1-\gamma}
\mbox{ \RM{ if $R \rightarrow 0 $ }},
\end{eqnarray}
where $H(x)$ is a step function, which is unity for $x>0$ and zero
otherwise, and
\begin{equation}\label{cn}
c_n = {2 \nu_0 b \sqrt{\pi} \Gamma(1+{n \over 2}) \over n \Gamma({1+n \over 2}) },~~~ n >0,
\end{equation}
and
\begin{equation}\label{i0}
I_0 =2 \nu_0 b \alpha B(\alpha(\beta-1), \alpha(1-\gamma)).
\end{equation}

For the potential $\Phi(r)$
\begin{eqnarray}\label{phiasm.1}
{\Phi(r) \over 4\pi \nu_0 b^2 } & \rightarrow & \alpha {b \over r} 
B(\alpha (3-\gamma ), \alpha ( \beta -3)) H(\beta-3) + {1 \over (3-\beta)(\beta-2)} ({b \over r})^{\beta-2} 
\mbox{ \RM{if $r\rightarrow +\infty$}}\\ \label{phiasm.2}
        & \rightarrow & \alpha 
B(\alpha (\beta -2), \alpha (2 -\gamma)) H(2-\gamma) +  {1 \over (3-\gamma)(\gamma-2)} ({b \over r})^{\gamma-2} 
\mbox{ \RM{if $ r \rightarrow 0$}} 
\end{eqnarray}
where $\beta>[2, \gamma]_{min}$, $0\leq \gamma<3$.  The zero point
of the potential is at infinitely large radius.  The depth of the
potential well $\Phi_0=\Phi(0)>0$ is infinite for models with a strong
cusp with $\gamma \geq 2$ (or if there is a central black hole), and
is finite for $\gamma<2$.

For the phase space density $f(Q)={d \over dQ} G(Q)$, we have
\begin{eqnarray} 
G(Q) &=& {1 \over 2\pi^2 \sqrt{2}} 
\int_{Q \ge \Phi(r)} {d \nu(r) \over dr} {1 \over \sqrt{Q-\Phi(r)}} dr \\\label{gasm.1}
     & \propto & q^{\beta_{2e}}
\mbox{ \RM{if $Q \rightarrow 0$}}\\ \label{gasm.2}
     & \propto & q^{\gamma_{2e}}
\mbox{ \RM{if $ Q \rightarrow \Phi(0) $}} ,
\end{eqnarray}
where $q$ is given in Eq.~\ref{qdef}, and 
\begin{equation}\label{beta2e}
\beta_{2e} = [ {\beta \over \beta-2 }- {1 \over 2}, \beta - {1 \over 2} ]_{max},
\end{equation}
and
\begin{eqnarray}\label{gamma2e}
\gamma_{2e} &= & {\gamma +2 \over 2|\gamma-2| } \mbox{ \RM{if $\gamma >0$  } } \\
     & = & [{1 \over 2 } (1 -{1 \over \alpha}),0]_{max} \mbox{ \RM{if $\gamma=0$.  } } 
\end{eqnarray}

\section{Other Approximate Models}

A disadvantage of the fit formulae purposed in the main text for the
volume density and the phase space density is that they do not follow
rigorously with the analytical asymptotic expressions at small radius.
Rather the parameters $\gamma$ and $\gamma_2$ are free fitting
parameters adjusted to fit all radii equally well.  This might be OK
for fitting observations with finite resolution at the center, but is
not satisfactory for theoretical modelling.

On the other hand, it is possible to devise other expressions which
have the expected analytical asymptotic behavior while still keeping
the residual small near the transition region.

The projected density $I(R)$ of the double-power-law
density $\nu(r)$ (cf. Eq.~\ref{nu}, and~\ref{IR}) satisfies the following
approximation,
\begin{equation}\label{irapprox}
I(R)  \approx  I_{in}(R) + I_{out}(R) ,
\end{equation}
where
\begin{equation}\label{in}
I_{in}(R)  =  c_{\gamma-1} 
({R \over b})^{1-\gamma} 
(1+({R \over b})^{1 \over \alpha})^{-1-\alpha(\beta-\gamma)},
\mbox{ \RM{ if $\gamma>1 $ }},
\end{equation}
and
\begin{equation}\label{out}
I_{out}(R)  =  c_{\beta-1} 
({R \over b})^{1-\gamma+ {1 \over \alpha} }
(1+({R \over b})^{1 \over \alpha})^{-1-\alpha(\beta-\gamma)},
\end{equation}
where $c_n$ is given in Eq.~\ref{cn}.

The approximation is devised so that $I_{in}(R)+I_{out}(R)$ is
rigorously $I(R)$ at asymptoticly big or small radius.
\begin{equation}
I(R) \rightarrow I_{in}(R) \rightarrow 
c_{\gamma-1} ({R \over b})^{1-\gamma} \gg I_{out}(R)
\mbox{ \RM{ if $R \rightarrow 0 $ }},
\end{equation}
and
\begin{equation}
I(R) \rightarrow I_{out}(R) \rightarrow 
c_{\beta-1} ({R \over b})^{1-\beta} \gg I_{in}(R)
\mbox{ \RM{if $R \rightarrow +\infty $ }}.
\end{equation}

The approximation is also found to be typically accurate within a
$10\%$ in the transition region.  This is qualitatively understandable
if ones notes the equality
\begin{equation}
\nu_{\alpha,\beta,\gamma} (r) = \nu_{\alpha,\beta+{1\over \alpha},\gamma} (r) 
+ \nu_{\alpha,\beta,\gamma-{1\over \alpha}} (r) ,
\end{equation}
where the subscripts specify the double-power-law slopes, and that the
projected density of $\nu_{\alpha,\beta-{1\over \alpha},\gamma} (r)$
and $\nu_{\alpha,\beta,\gamma-{1\over \alpha}} (r)$ is roughly
$I_{in}(R)$ and $I_{out}(R)$ respectively.

The above suggests that it is worthwhile to fit the real photometric
data with $I_{in}(R)+I_{out}(R)$ instead of $\mu(R)$, because one
obtains simple accurate expressions for the projected density and the
volume density simultaneously.  But note that if $\gamma \le 1$,
namely, if the projected profile has a finite core, the expression for
$I_{in}(R)$ is undefined (because $c_{\gamma-1}$ is undefined, see
Eq.~\ref{cn}).

With similar techniques, one can also work out simple approximation to 
the phase space density corresponding to the double-power-law.  The idea is
to write 
\begin{equation}
f(Q) = f_{in}(Q) + f_{out} (Q),
\end{equation}
so that $f_{in}(Q)$ is approximately consistent with
$\nu_{\alpha,\beta-{1\over \alpha},\gamma} (r)$ and $f_{out}(Q)$ with
$\nu_{\alpha,\beta,\gamma-{1\over \alpha}} (r)$.  The results (not
given here) are somewhat tedious depending on the range of $\gamma$
and $\beta$.

\section{Line profile expressed as a 1D integral}

Here we show that the line profile can be reduced to a 1D integral for
the models.  With no loss of simplicity we will derive the equations
in the slightly general context of a Ospikov-Merritt type anisotropic
model, where the distribution function
\begin{equation}
f(E,J)=f(Q_a), \,\, \mbox{ \RM{ $Q_a \equiv -E-{1 \over 2} \eta
J^2$}}
\end{equation}
where $\eta \equiv {1 \over r_a^2}$ has the dimension of inverse
squared distance, and $r_a$ is an anisotropy radius.  If $\eta >0$ ,
then beyond $r_a$ most orbits are radial.  The model reduces to
isotropic with $f=f(E)$ in the special case that $\eta=0$.

Generally
\begin{equation}
Q_a=\Phi(r)-{1 \over 2}(v_x^2+v_y^2+v_z^2+\eta J^2),
\end{equation}
and
\begin{eqnarray}
J^2 & = & (x v_y -y v_x)^2 +(x v_z- z v_x)^2 +(z v_y - y v_z)^2,\\
   & = & v_x^2 (y^2+z^2) + v_y^2 (x^2+z^2) + v_z^2 (x^2+y^2) 
  -2(xyv_xv_y + yzv_yv_z + zxv_zv_x).
\end{eqnarray}
Without loss of generality, we can set
\begin{equation}
x=R, \,\, y=0
\end{equation}
It then follows that
\begin{equation}
Q_a= \Phi(r)-{1 \over 2}(a_R v_z^2 + a_z v^2_x  - 2a_{Rz} v_zv_x + a_r v^2_y)
\end{equation}
where
\begin{equation}
a_R= (1+\eta R^2), \,\, 
a_z= (1+\eta z^2), \,\, 
a_r= (1+\eta z^2+ \eta R^2), \,\, 
a_{Rz}= \eta Rz.
\end{equation}
With a change of variables
\begin{equation}
v_y = v'_y,\,\, v_x = v'_x + u,\,\, 
u={a_{Rz}v_z \over a_z},
\end{equation}
we have
\begin{equation}
dv_x dv_y = dv'_x dv'_y.
\end{equation}
and
\begin{equation}
Q_a= \Phi(r)-{1 \over 2}(v_z^2 a_{rz} +v'^2_x a_z + v'^2_y a_r ),
\end{equation}
where
\begin{equation}
a_{rz}= (a_R - {a^2_{Rz} \over a_z} ) = {1+\eta (R^2+z^2) \over 1+ \eta z^2}.
\end{equation}
With a further transformation of coordinates, one finds that
\begin{equation}
\int\int dv_x dv_y f(Q_a) = \int_0^{Q_a} \int_0^{2 \pi} d\theta d Q_a f(Q_a) {1 \over \sqrt{a_r a_z} }.
\end{equation}

If one can devise a function $G_a(Q_a)$ as an elementary function of
$Q_a$ and specify the phase space density $f(Q_a)$ by
\begin{equation}
f(Q_a) = {d \over dQ_a} G_a(Q_a),
\end{equation}
then the 3D integral for the line profile is reduced to a 1D integral,
\begin{eqnarray}
P(R, v_z) & = & \int^\infty_{-\infty} dz \int\int dv_x dv_y f(Q_a)
          \\\label{profileani} 
          & = & 4\pi \int_0^\infty dz {G_a(Q_a)
          \over (1+ \eta (R^2+z^2) )^{1/2} (1+\eta z^2)^{1/2} },
\end{eqnarray}
where
\begin{equation}
Q_a=\Phi(\sqrt{R^2+z^2})- {v_z^2 \over 2} {1+\eta (R^2+z^2) \over 1+
         \eta z^2} .
\end{equation}
When $\eta=0$ and $G_a(Q_a)=G(Q)$, the above reduces to 
Equation~\ref{profile} of isotropic models.


\begin{deluxetable}{rrrrrrrrrrrrr}
\tablewidth{33pc}
\tablecaption
 {Fitting the surface density and volume density at points
 ${R_i \over B}={r_i \over B}$ for $i=1,...,N_R=N_r=13$.}
 \tablehead{\colhead{i=1} &\colhead{2} &\colhead{3} &\colhead{4} &\colhead{5}
&\colhead{6} &\colhead{7} &\colhead{8} &\colhead{9} &\colhead{10} &\colhead{11}
&\colhead{12} &\colhead{13} }
\startdata
0.01 & 0.02 & 0.05 & 0.1 & 0.2 & 0.5 & 1 & 2 & 5 & 10 & 20 & 50 & 100 \nl
\tablecomments{
$B$ is the break radius of the surface brightness profile.
}
\enddata
\end{deluxetable}

\begin{deluxetable}{rrrrrrrrrr}
\tablewidth{33pc}
\tablecaption{
fit the line profiles at the following velocities ${v_j \over V_{esc}(R)}$ 
for $j=1,...,N_v=10$.}
\tablehead{\colhead{j=1} &\colhead{2} &\colhead{3} &\colhead{4} &\colhead{5}
&\colhead{6} &\colhead{7} &\colhead{8} &\colhead{9} &\colhead{10} }
\startdata
 0 & 0.1 & 0.2 & 0.3 & 0.4 & 0.5 & 0.6 & 0.7 & 0.8 & 0.9 \nl
\tablecomments{
$V_{esc}(R)$ is the escape velocity at radius $R$.  The fit is done
for radii $R=R_i$ with $i=4,...,N_R=13$ as in Table 1.}
\enddata
\end{deluxetable}

\begin{deluxetable}{rrrrrrrrrrrrr}
\tablewidth{42pc}
\tablecaption
{The parameters for 
models shown in Fig. 1, Fig. 3 and Fig. 5:
the surface brightness profile $(\alpha_1, \beta_1, \gamma_1)$, 
the volume density model $(\alpha, \beta, \gamma, b, \nu_0)$ and 
its rms residual, the phase space density model
$(\alpha_2, \beta_2, \gamma_2, Q_b, f_0)$ and its rms residual. }
 \tablehead{
\colhead{$\alpha_1\_\beta_1\_\gamma_1$} & 
 \colhead{$\alpha$} & \colhead{$\beta$} & \colhead{$\gamma $} & 
 \colhead{$b$} & \colhead{$\nu_0$} & \colhead{rms} &
 \colhead{$\alpha_2$} & \colhead{$\beta_2$} & \colhead{$\gamma_2$} & 
\colhead{$Q_b$} & \colhead{$\ln f_0$} & \colhead{rms} }
 \startdata
 $0.5\_1.8\_0.0$&0.50&2.80&0.00&1.00&0.47&1.E-03&0.64&3.50&0.16&27.80&-7.40&7.E-02\nl
 $0.5\_1.8\_0.5$&0.50&2.80&1.36&1.36&0.20&3.E-02&0.25&3.50&2.34&18.90&-7.99&2.E-01\nl
 $0.5\_1.8\_1.0$&0.50&2.80&1.97&1.40&0.19&1.E-02&0.67&3.50&9.99&14.47&-8.66&2.E-01\nl
 $0.5\_1.8\_1.5$&0.50&2.80&2.49&1.37&0.20&3.E-03&1.87&3.50&5.10&30.13&-8.30&6.E-03\nl
\tableline
 $0.5\_1.2\_0.5$&0.63&2.20&1.36&1.26&0.22&3.E-02&1.12&11.00&2.27&13.79&-12.04&1.E-01\nl
 $0.5\_1.8\_0.5$&0.50&2.80&1.36&1.36&0.20&3.E-02&0.25&3.50&2.34&18.90&-7.99&2.E-01\nl
 $0.5\_2.4\_0.5$&0.50&3.40&1.33&1.29&0.24&3.E-02&0.25&3.40&2.17&9.49&-7.31&2.E-01\nl
 $0.5\_3.0\_0.5$&0.50&4.00&1.31&1.23&0.28&4.E-02&0.37&4.00&2.06&3.00&-7.73&7.E-02\nl
\tableline
 $0.5\_1.4\_0.0$&0.51&2.40&0.00&1.00&0.40&4.E-03&1.29&6.00&0.00&23.56&-6.54&2.E-02\nl
 $1.0\_1.4\_0.0$&1.05&2.40&0.26&0.85&0.60&3.E-03&1.27&6.00&0.55&18.72&-7.67&6.E-03\nl
 $1.5\_1.4\_0.0$&1.57&2.40&0.44&0.79&0.73&2.E-03&1.31&6.00&0.83&17.18&-7.98&2.E-02\nl
 $2.0\_1.4\_0.0$&2.00&2.40&0.64&0.94&0.47&5.E-03&1.29&6.00&1.11&14.93&-8.40&2.E-02\nl
\tablecomments{
The break radius $B$ and the normalization $\mu_0$ are set to unity.  
 $(\alpha_1, \beta_1, \gamma_1)$ and $(\alpha, \beta, \gamma)$
 describe the width of the transition region,
the slope of the outer power-law and inner power-law for the surface 
 brightness profile and for the volume density profile.   
$b$ and $\nu_0$ are the scales for the radius and the volume density.}
\enddata
\end{deluxetable}

\vfill\eject

\begin{deluxetable}{rrrrrrrrrrrrl}
\tablewidth{42pc}
\tablecaption {For each galaxy the table gives surface bright profile
parameters $(\alpha_1, \beta_1, \gamma_1)$$\ddagger$ found by Byun et
al. (1996) by fitting observation, the five volume density model
parameters $(\alpha, \beta, \gamma, b, \nu_0)$, the five phase space
density model parameters $(\alpha_2, \beta_2, \gamma_2, Q_b, f_0)$
found here.  The last column shows the rms residuals (multiplied by
100) of Byun et al. fit, of our volume density model, of our phase
space density model respectively.}  
\tablehead{
\colhead{Galaxy} & 
\colhead{$\alpha_1\_\beta_1\_\gamma_1$} & 
\colhead{$\alpha$} & \colhead{$\beta$} & \colhead{$\gamma $} & 
\colhead{$b$} & \colhead{$\nu_0$} &
\colhead{$\alpha_2$} & \colhead{$\beta_2$} & \colhead{$\gamma_2$} & 
\colhead{$Q_b$} & \colhead{$\ln f_0$} &
\colhead{RMS} }
\startdata
N~596&$1.3\_2.0\_0.6$&1.29&2.97&1.47&1.32&0.22&0.79&2.57&3.25&2E-3&-25.6&3:0:2\nl
N~720&$0.4\_1.7\_0.1$&0.37&2.66&0.44&1.22&0.26&0.75&3.53&0.64&17.6&-8.17&1:0:0\nl
N1172&$0.7\_1.6\_1.0$&0.64&2.64&1.98&1.40&0.18&0.47&3.63&9.99&13.0&-8.54&3:0:4\nl
N1399&$0.7\_1.7\_0.1$&0.63&2.68&0.52&1.15&0.31&0.70&3.43&0.77&18.3&-8.07&1:0:0\nl
N1400&$0.7\_1.3\_0.0$&0.74&2.32&0.13&0.92&0.47&1.31&6.77&0.32&20.0&-7.45&3:0:0\nl
N1600&$0.8\_2.2\_0.0$&0.81&3.18&0.20&0.94&0.64&0.40&2.68&0.54&16.8&-6.90&2:0:4\nl
N1700&$1.1\_1.3\_0.0$&1.14&2.30&0.35&0.88&0.52&1.34&7.10&0.64&17.7&-8.27&2:0:0\nl
N2832&$0.5\_1.4\_0.0$&0.54&2.40&0.23&1.07&0.34&1.13&5.51&0.40&19.2&-7.80&1:0:0\nl
N3115&$0.7\_1.4\_0.8$&0.67&2.43&1.75&1.46&0.16&0.35&5.16&4.78&0.92&-20.4&2:0:4\nl
N3377&$0.5\_1.3\_0.3$&0.41&2.33&1.12&1.55&0.14&0.81&6.52&1.63&13.6&-10.7&3:0:0\nl
N3379&$0.6\_1.4\_0.2$&0.55&2.43&0.89&1.40&0.18&0.90&5.13&1.20&13.9&-9.35&2:0:0\nl
N3608&$1.0\_1.3\_0.0$&0.98&2.33&0.27&0.89&0.51&1.31&6.56&0.53&18.3&-7.94&2:0:0\nl
N4168&$1.1\_1.5\_0.1$&1.04&2.50&0.80&1.17&0.28&1.00&4.50&1.14&15.9&-8.35&2:0:0\nl
N4365&$0.5\_1.3\_0.1$&0.40&2.27&0.81&1.44&0.16&0.93&7.91&1.06&17.8&-10.2&2:0:1\nl
N4464&$0.6\_1.7\_0.9$&0.55&2.68&1.85&1.45&0.17&0.86&3.43&9.99&44.9&-5.75&2:0:1\nl
N4551&$0.3\_1.2\_0.8$&0.30&2.23&1.77&1.52&0.14&0.16&9.08&5.77&11.1&-16.4&3:0:2\nl
N4552&$0.7\_1.3\_0.0$&0.70&2.30&0.10&0.93&0.45&1.33&7.23&0.27&20.5&-7.34&4:0:0\nl
N4621&$5.3\_1.7\_0.5$&5.44&2.71&1.43&1.04&0.44&0.59&3.32&5.52&8E-3&-24.6&3:0:0\nl
N4636&$0.6\_1.3\_0.1$&0.55&2.33&0.75&1.34&0.19&0.98&6.58&1.00&16.2&-9.52&1:0:0\nl
N4649&$0.5\_1.3\_0.2$&0.41&2.30&0.82&1.44&0.17&0.92&7.17&1.08&16.5&-10.0&3:0:1\nl
N4874&$0.4\_1.4\_0.1$&0.33&2.37&0.76&1.42&0.17&0.93&5.96&0.98&14.9&-9.45&1:0:1\nl
N4881&$0.6\_1.4\_0.8$&0.53&2.36&1.71&1.50&0.15&0.58&6.09&4.71&1.91&-19.1&3:0:2\nl
N4889&$0.4\_1.3\_0.0$&0.33&2.35&0.36&1.23&0.24&1.07&6.25&0.51&17.9&-8.42&1:0:1\nl
N5813&$0.5\_1.3\_0.1$&0.41&2.33&0.54&1.29&0.22&1.01&6.64&0.73&17.4&-9.06&2:0:1\nl
N5845&$0.8\_2.7\_0.5$&0.74&3.74&1.36&1.24&0.26&0.27&3.24&2.32&3.21&-7.40&8:0:0\nl
\tablecomments{The $\alpha_1$ here is the inverse of the 
$\alpha$ used by Byun et al. (1996).  Dimensional quantities are
normalized with $B$ and $\mu_0$.
$(\alpha_1, \beta_1, \gamma_1)$ and $(\alpha, \beta, \gamma)$
 describe the width of the transition region,
the slope of the outer power-law and inner power-law for the surface 
 brightness profile and for the volume density profile.   
$b$ and $\nu_0$ are the scales for the radius and the volume density.}
\enddata
\end{deluxetable}

\vfill\eject

\begin{deluxetable}{rrrrrrrrrrrrl}
\tablewidth{42pc}
\tablecaption
{The predicted line-of-sight velocity dispersions $\sigma(R)$ 
and kurtosis parameter $c_4(R)$ at radii $R/B=0.1,0.2,0.5,1,2,5,10,20,50,100$.}
\tablehead{
\colhead{Galaxy} 
& \colhead{$\sigma:0.1$} &\colhead{$0.2$} &\colhead{$0.5$} 
& \colhead{$1$} & \colhead{$2$} &\colhead{$5$} 
& \colhead{$10$} &\colhead{$100$} 
&\colhead{$c_4:0.1$} &\colhead{$0.2$} &\colhead{$0.5$} &\colhead{$|c_4(\ge 1)|$} }
\startdata
N~596&0.67&0.72&0.74&0.72&0.67&0.57&0.49&0.23&0.02&0.01&0.00&0.02\nl
N~720&0.77&0.84&0.93&0.97&0.95&0.85&0.75&0.42&0.18&0.11&0.04&0.02\nl
N1172&1.41&1.36&1.28&1.20&1.11&0.96&0.82&0.45&-0.01&-0.01&-0.01&0.02\nl
N1399&0.67&0.74&0.83&0.86&0.85&0.78&0.70&0.40&0.17&0.09&0.03&0.02\nl
N1400&0.74&0.78&0.86&0.92&0.97&0.98&0.95&0.75&0.13&0.10&0.05&0.02\nl
N1600&0.61&0.64&0.65&0.63&0.62&0.55&0.46&0.20&0.08&0.03&-0.01&0.02\nl
N1700&0.55&0.62&0.72&0.79&0.85&0.89&0.89&0.75&0.14&0.10&0.05&0.03\nl
N2832&0.79&0.84&0.92&0.97&1.00&0.97&0.91&0.65&0.15&0.10&0.04&0.02\nl
N3115&1.10&1.13&1.14&1.13&1.09&1.01&0.92&0.62&0.01&0.01&0.00&0.02\nl
N3377&0.69&0.81&0.96&1.05&1.09&1.06&0.99&0.72&0.11&0.08&0.05&0.02\nl
N3379&0.66&0.76&0.89&0.97&1.00&0.96&0.88&0.60&0.15&0.10&0.05&0.02\nl
N3608&0.61&0.67&0.77&0.83&0.89&0.91&0.90&0.72&0.14&0.10&0.05&0.03\nl
N4168&0.55&0.63&0.73&0.78&0.81&0.79&0.75&0.52&0.12&0.08&0.03&0.02\nl
N4365&0.68&0.81&0.95&1.05&1.11&1.10&1.05&0.81&0.17&0.12&0.06&0.03\nl
N4464&1.20&1.21&1.19&1.15&1.07&0.91&0.78&0.42&0.00&0.00&-0.00&0.02\nl
N4551&1.11&1.16&1.20&1.23&1.23&1.20&1.15&0.87&0.01&0.01&0.01&0.02\nl
N4552&0.78&0.81&0.88&0.94&0.99&1.01&0.98&0.79&0.12&0.09&0.05&0.02\nl
N4621&0.15&0.15&0.15&0.15&0.15&0.14&0.13&0.10&0.00&0.00&-0.00&0.02\nl
N4636&0.67&0.77&0.91&0.99&1.04&1.02&0.97&0.73&0.16&0.12&0.06&0.02\nl
N4649&0.68&0.80&0.95&1.04&1.09&1.07&1.01&0.77&0.16&0.12&0.06&0.02\nl
N4874&0.71&0.81&0.95&1.04&1.08&1.04&0.97&0.69&0.19&0.13&0.06&0.02\nl
N4881&1.05&1.09&1.14&1.15&1.14&1.07&1.00&0.72&0.01&0.01&0.00&0.02\nl
N4889&0.82&0.88&0.98&1.06&1.09&1.05&0.98&0.72&0.19&0.13&0.06&0.02\nl
N5813&0.75&0.83&0.95&1.04&1.08&1.05&0.99&0.74&0.20&0.14&0.06&0.02\nl
N5845&0.74&0.79&0.81&0.75&0.63&0.46&0.35&0.12&0.03&0.02&-0.00&0.02\nl
\tablecomments{  The last
column is the maximum of $|c_4(R)|$ outside the core $R \ge 1$.  
The dispersion $\sigma$ is normalized with 
$(G\mu_0 B)^{1 \over 2}$, where $\mu_0$ and $B$
are defined in Eq.(1) and $G$ is the gravitational constant.  }
\enddata
\end{deluxetable}

\vfill\eject

\begin{figure}[] \vbox to5.5in{\rule{0pt}{5.5in}}
\special{voffset=-60 hoffset=30 vscale=70 hscale=70 psfile=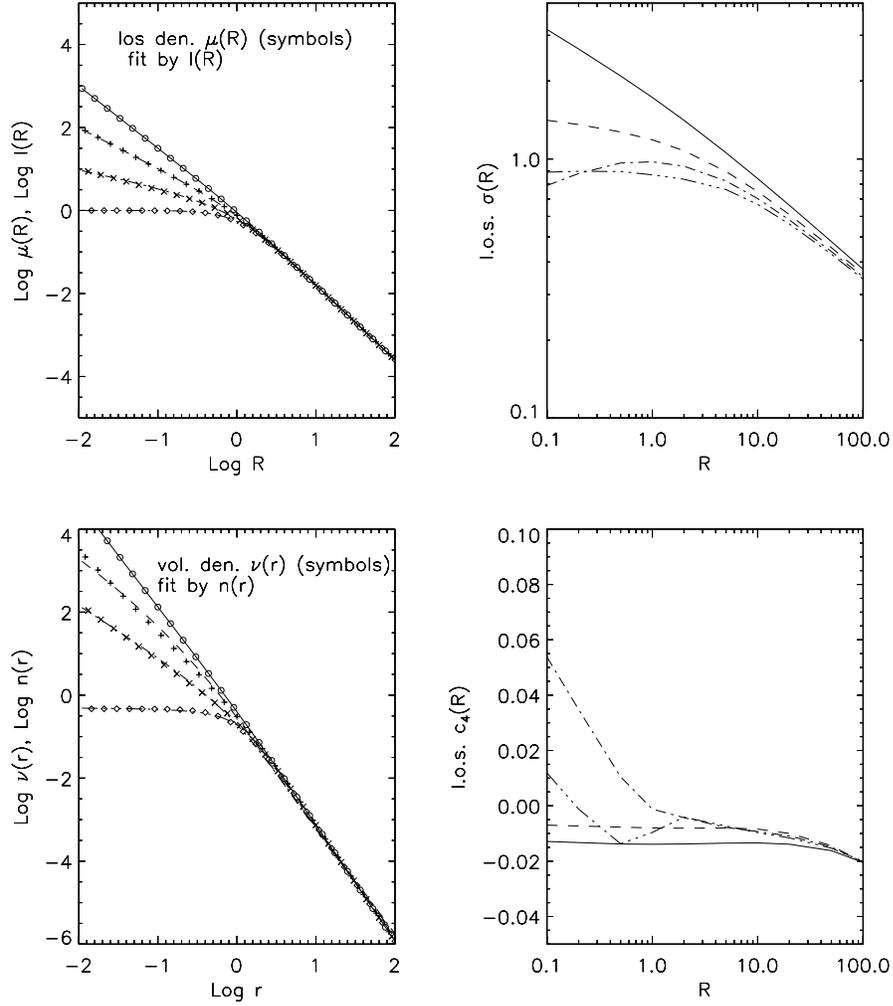} 
\caption{shows a) several fits to the surface brightness profiles $\mu(R)$
by integrating the double-power-law model volume density models, b)
several fits to the volume density profiles $\nu(r)$ by integrating
the phase space density models, c) the radial run of the predicted
line-of-sight dispersion $\sigma(R)$ and d) kurtosis $c_4(R)$ for the
isotropic double-power-law dynamical model.  The four models have
$(\alpha_1, \beta_1, \gamma_1)=(0.5,1.8,\gamma_1)$ and $\gamma_1=0,
0.5,1, 1.5$, which are shown by the fit of four different type of
lines (solid, dashed, dash dot, and dash dot dot dot lines) to four
different symbols (circles, pluses, crosses, and diamonds)
respectively. }\label{fig1.ps}
\end{figure}

\begin{figure}[] \vbox to5.5in{\rule{0pt}{5.5in}}
\special{voffset=-60 hoffset=30 vscale=70 hscale=70 psfile=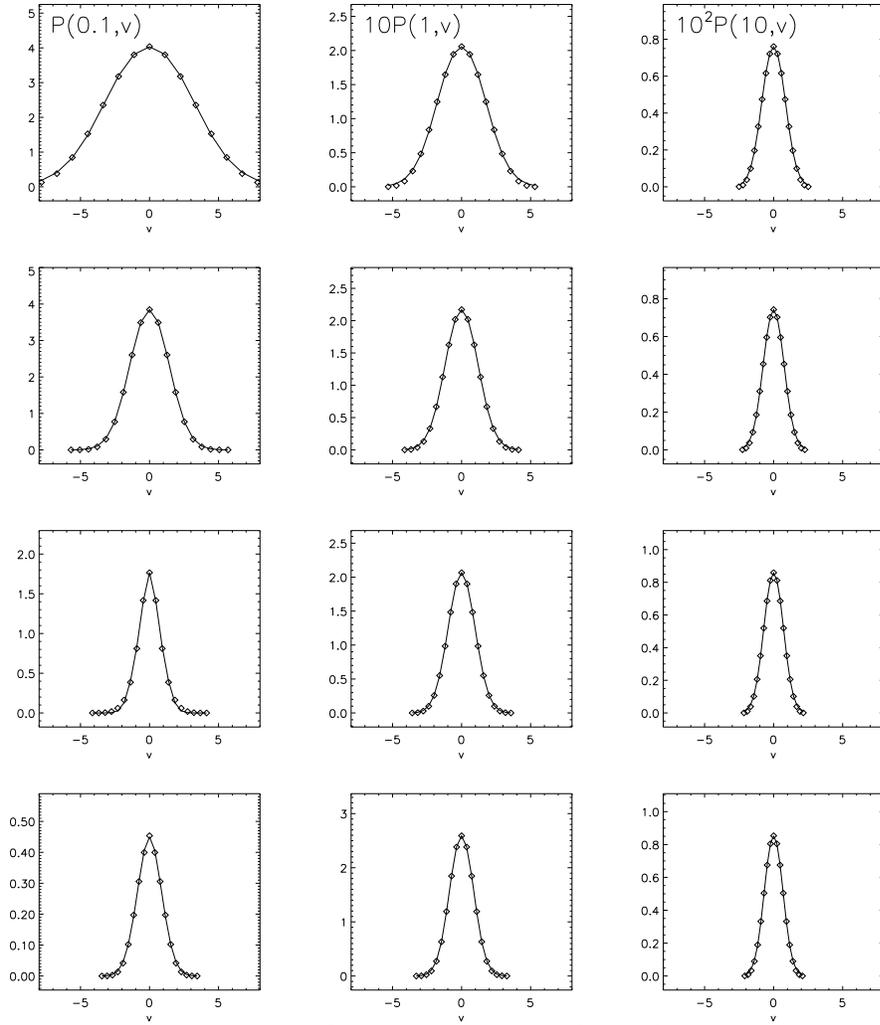} 
\caption{from top to down shows the fits to the line profiles of the 
models shown in Fig.1 with the parameterization by Zhao and
Prada (1996).  The three columns correspond to profiles at projected
radius $R=0.1,1,10$ from left to right.  All models have nearly
Gaussian profiles.}\label{fig2.ps}
\end{figure}

\begin{figure}[] \vbox to5.5in{\rule{0pt}{5.5in}}
\special{voffset=-60 hoffset=30 vscale=70 hscale=70 psfile=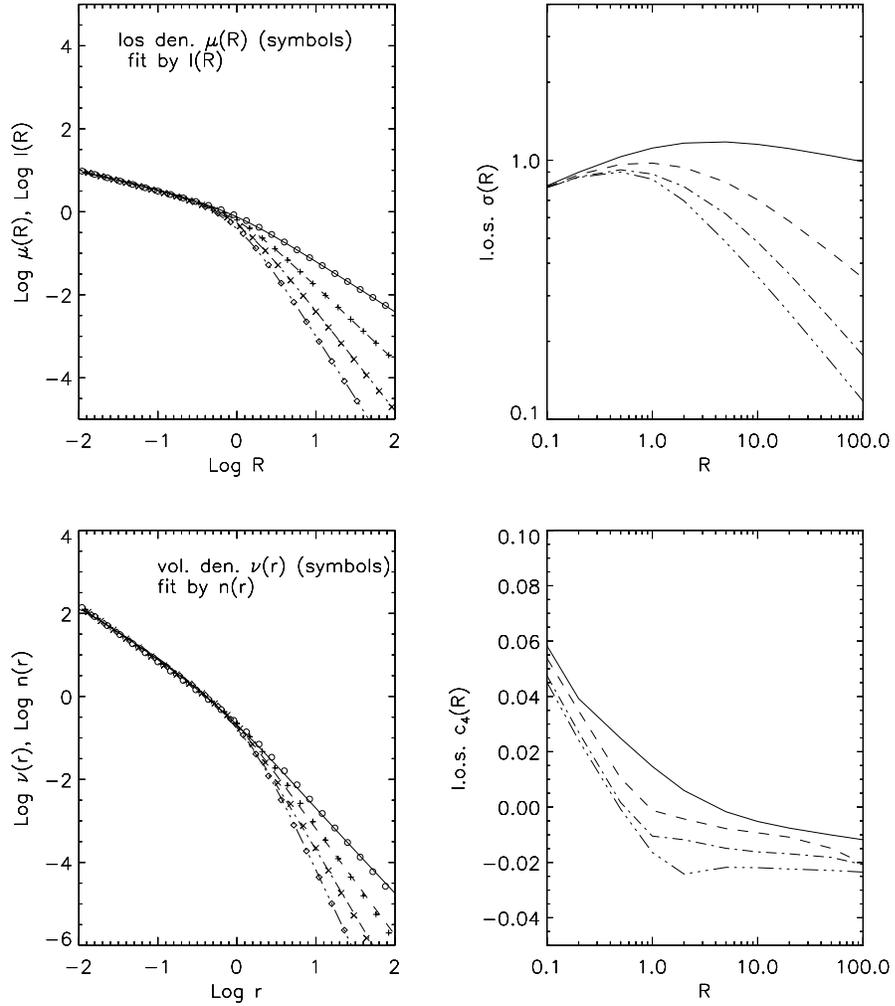} 
\caption{similar to Fig.1 except that models have $(\alpha_1, \beta_1, 
\gamma_1)=(0.5,\beta_1,0.5)$ and $\beta_1=1.2, 1.8, 2.4, 3$. }
\label{fig3.ps}
\end{figure}

\begin{figure}[] \vbox to5.5in{\rule{0pt}{5.5in}}
\special{voffset=-60 hoffset=30 vscale=70 hscale=70 psfile=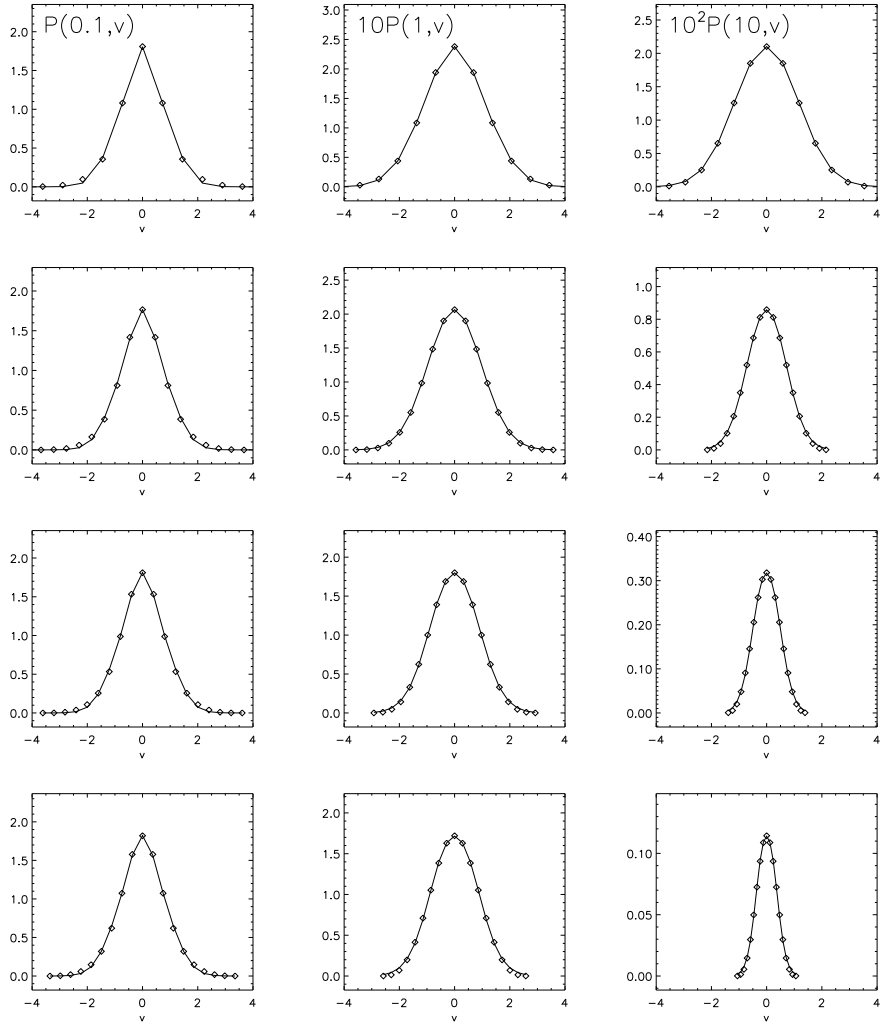} 
\caption{similar to Fig.2 except that the models are those in Fig.3.}
\label{fig4.ps}
\end{figure}

\begin{figure}[] \vbox to5.5in{\rule{0pt}{5.5in}}
\special{voffset=-60 hoffset=30 vscale=70 hscale=70 psfile=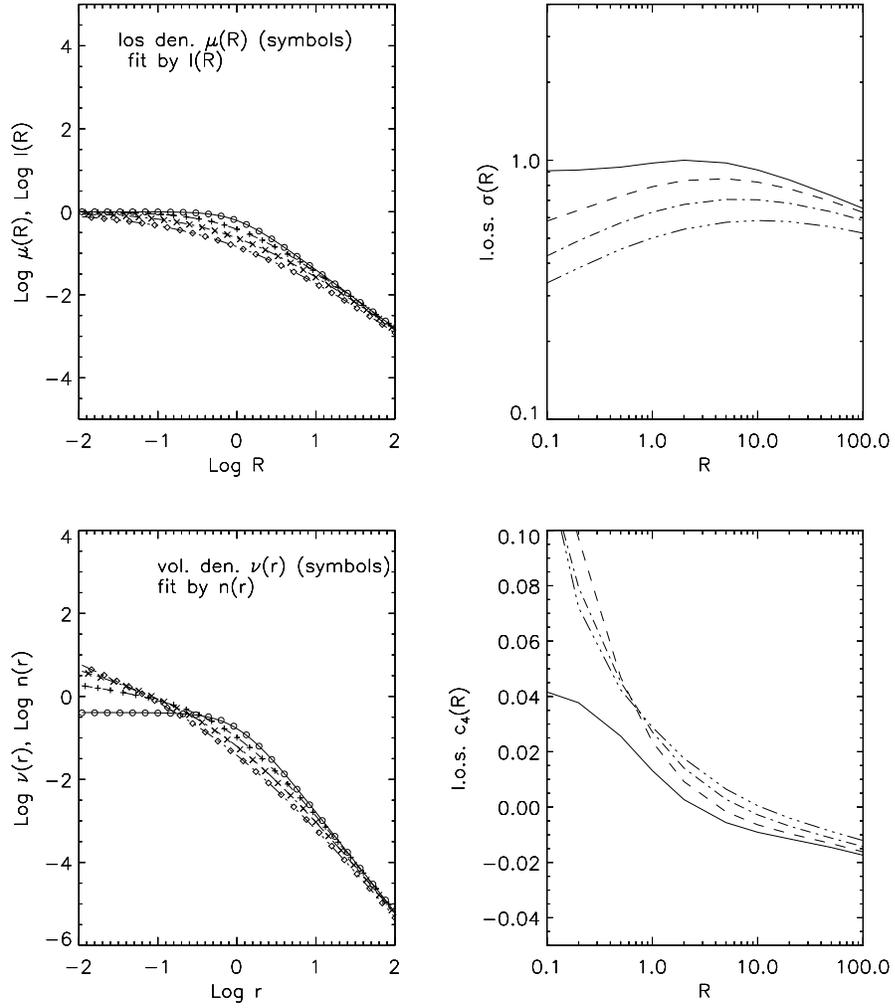} 
\caption{similar to Fig.1 except that models have $(\alpha_1, \beta_1, 
\gamma_1)=(\alpha_1,1.4,0)$ and $\alpha_1=0.5,1,1.5,2.$. }
\label{fig5.ps}
\end{figure}

\begin{figure}[] \vbox to5.5in{\rule{0pt}{5.5in}}
\special{voffset=-60 hoffset=30 vscale=70 hscale=70 psfile=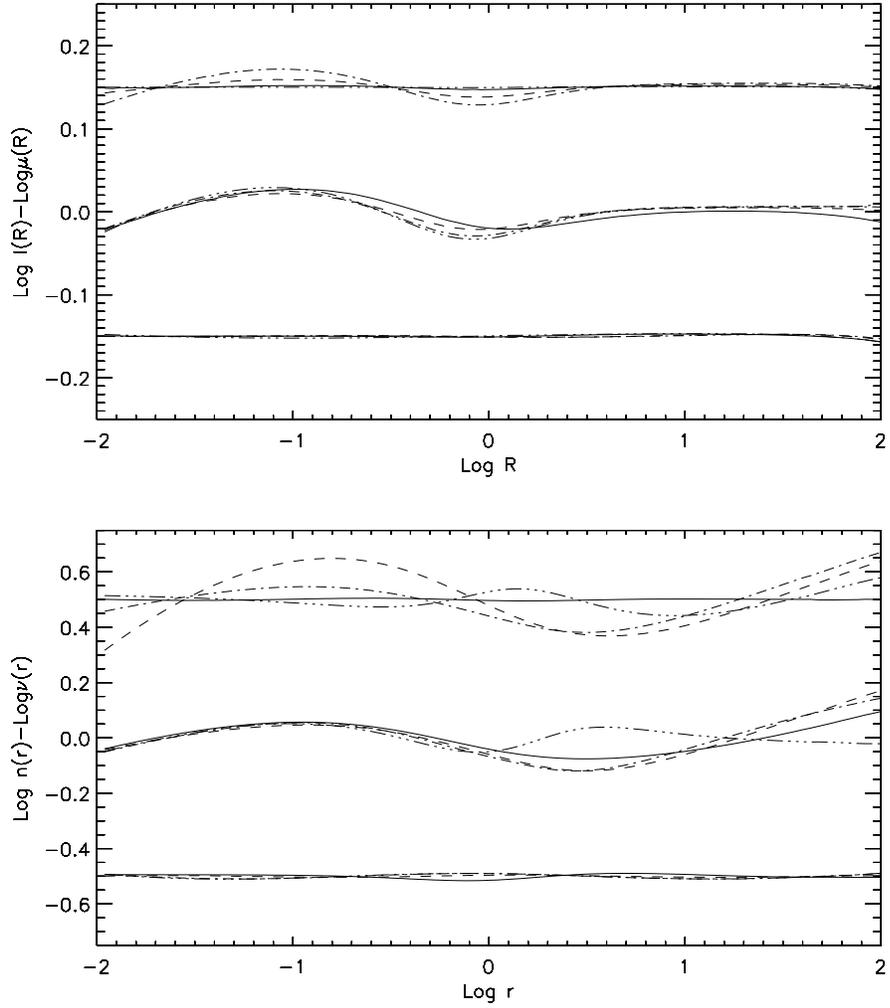} 
\caption{from top to down shows the residual as a run of radius
 of the models in Table 3 for the surface density (upper panel) and
the volume density (lower panel).  For clarity, the residual of three
sets of models has been offset by $\pm 0.15$ for surface density and
$\pm 0.5$ for volume density.  }\label{fig6.ps}
\end{figure}

\end{document}